\documentstyle[12pt]{article}

\def\di{\displaystyle}
\def\&{&\di}
\def\bg{\begin{eqnarray}\begin{array}{rcl}\displaystyle}
\def\eg{\end{array} &\di    &\di   \end{eqnarray}}
\def\bm#1{\begin{eqnarray}\begin{array}{#1}\di}
\def\bmo#1{\begin{eqnarray*}\begin{array}{#1}\di}
\def\eg{\end{array} &\di    &\di   \end{eqnarray}}
\def\bgo{\begin{eqnarray*}\begin{array}{rcl}\displaystyle}
\def\ego{\end{array} &\di    &\di \nonumber  \end{eqnarray*}}

\def\btensor#1#2{\renew\left#1\begin{array}{#2}\di}
\def\etensor#1{\end{array}\right#1}

\def\eq#1{eq.~(\ref{#1})}

\def\d{{\mbox d}}

\def\Tr{\mbox{Tr}}

\def\T{{\mbox T}}

\def\id{1\!\mbox{l}}

\def\p{\partial\llap{/}}

\def\A{A\!\llap{/}}

\def\xv{\mbox{\boldmath$x$}}

\def\yv{\mbox{\boldmath$y$}}
\def\zv{\mbox{\boldmath$z$}}

\def\del{\mbox{\boldmath$\delta$}}

\def\CA{{\cal A}}

\def\CO{{\cal O}}

\def\llangle{\left\langle}
\def\rrangle{\right\rangle}

\date{\today}

\def\rene{\renewcommand{\arraystretch}{1.9}}
\def\renew{\renewcommand{\arraystretch}{1}}
\rene

\newcommand{\mysection}[1]{\section{#1}\setcounter{figure}{0}
\setcounter{table}{0}\setcounter{equation}{0}}

\begin{document}

\begin{titlepage}

\parindent=12pt
\baselineskip=20pt
\textwidth 15 truecm
\vsize=23 truecm
\hoffset=0.7 truecm

\begin{flushright}
   FSUJ-TPI-18/96 \\
      \end{flushright}
\par
\vskip .5 truecm
\large \centerline{\bf The Gauss Law Operator Algebra}
\large \centerline{\bf and Double Commutators in Chiral Gauge Theories} 
\par
\vskip 1 truecm
\normalsize
\begin{center}
 {\bf J.~M.~Pawlowski}\footnote{e--mail:
  pawlowski@tpi.uni-jena.de}\\
\it{Theor.--Phys. Institut, Universit\"at Jena\\ 
Fr\"obelstieg 1\\
D--07743 Jena\\
Germany}
\end{center}
 \par
\vskip 2 truecm
\normalsize
\begin{abstract} 
We calculate within an algebraic Bjorken--Johnson--Low (BJL) method  
anomalous Schwinger terms of fermionic currents and the Gauss law
operator in chiral gauge theories. The current algebra is 
known to violate the Jacobi identity in an iterative
computation. Our method takes the subtleties of the 
equal--time limit into account and leads to an algebra that fulfills the
Jacobi identity. The non-iterative terms appearing in the double
commutators can be traced back directly to the projective
representation of the gauge group.  
\end{abstract}
       
\vfill
       
\end{titlepage}

              
\mysection{ Introduction}
Chiral gauge theories suffer after fermionic quantization 
from an anomalous breaking of gauge invariance due to the chiral
anomaly \cite{review}. The chiral anomaly is directly
related (via cohomological descent equations \cite{review}) to the
anomalous Schwinger term in the algebra of the Gauss law operator $G$ in these
theories \cite{fadeev}. The Gauss law operator consists of two parts
$G=G_A+G_\psi$, where $G_A$ generates (time-independent) 
gauge transformations on the gauge field and $G_\psi=J^0$ acts on fermions.
Here $J^0$ is the zero component of the (consistent) fermionic
current. The algebra of $G$ has been studied in many ways 
\cite{fadeev}-\cite{semenoff}, since it is 
connected to the question of consistency of quantized chiral gauge
theories. Whereas the cohomological prediction has been verified by the
results, the algebras of $G_{A/\psi}$ do not
coincide in general. Moreover, it is well known that an iterative 
calculation of double commutators
containing $G_{A/\psi}$ leads to a violation of the Jacobi identity. 
Double commutators of
fermionic currents obtained within an iterative computation 
do not fulfill a consistency condition, which relates them to the anomaly \cite{rothe}. 
If one quantizes the gauge field, the gauge field part $G_A$ is
formally given by the covariant derivative of the chromo-electric
field $E^a$ (\cite{jo},\cite{banerjee},\cite{semenoff}). In these
calculations an anomalous Schwinger term occurs in the 
commutator $[E^a,E^b]$. The double commutator
$\Bigl[E^a,[E^b,E^c]\Bigr]$ obtained in an
iterative computation 
violates the Jacobi identity. Therefore it is not clear whether the 
identification $G_A= -D\cdot E$ is correct. An explanation 
of these facts would give some insight to the 
structure of chiral gauge theories.  
       
 Since a violation of the Jacobi identity
also takes place in the case of free fermionic
axial and vector currents, one can study the free theory as a toy
model. The Schwinger terms in the algebra of free currents 
\cite{lev},\cite{double1} have been shown to be operator valued. 
Considerations concerning the commutator algebra of composite
operators in chiral gauge theories
\cite{doc} indicate that this is also true for the Schwinger terms
in the algebra of $G_{A/\psi}$. 
         
 In \cite{rothe} 
a BJL-type prescription \cite{bjl} was presented for the 
calculation of double commutators, which respects
the Jacobi identity and fulfills the consistency condition mentioned
before. 
Nevertheless it was not clear whether this prescription properly defines a double
commutator, since in the free theory the prescription fails to give the
correct result \cite{double1}. 
                
 In the present paper we give an algebraic BJL-type derivation 
of commutators and 
double commutators of $G_{A/\psi}$ based only on consistency. This is
related to the definition of a suitable regularization and
renormalization of the time evolution operator $U$ of the theory, and
shows up in a non-trivial renormalization factor of $U$ (e.g.
\cite{mickel},\cite{langmann}). The derivation gives a proper
definition of commutators and double commutators in chiral gauge
theories. Since little is known about a
consistent quantization of the gauge field in anomalous gauge
theories, we treat the gauge field as an external field with
$A(t=-\infty)=0$. This defines an
adiabatic solution of the theory \cite{semenoff}. We show that
the Schwinger terms of commutators can be derived algebraically from
the anomaly without further assumptions and without calculating
diagrams.The present approach also provides a simple explanation for the
discrepancies in the algebra of $G_{A/\psi}$ in the literature. 
The same method is applied to the calculation of
double commutators. As in the case of commutators we can 
derive all Schwinger terms from the anomaly only. 
The Jacobi identity is fulfilled non-trivially for the
algebra of $G_{A/\psi}$. The result confirms the validity of the
BJL-type prescription given in \cite{rothe} for the double
commutators of $G_{A/\psi}$. It also indicates, that this scheme should
be valid in general.  
                     
 In the second section we discuss the properties of the
BJL-limit  and give a definition of the Gauss law operator $G$ at
arbitrary times as the time evolution of $G(t=-\infty)$. 
The projective representation of the gauge group (on the Fock space) 
shows up in the 
properties of the time evolution operator. The following two sections
are dedicated to the derivation of equal-time
commutators and double commutators only using the anomaly equation. 
The last section summarizes the results.

\mysection{The Gauss Law Operator}\label{gauss}
In the following the gauge field is treated as an external field. The
fermionic action of a chiral gauge theory is 
\bm{rclcrcl}
\& S[\bar\psi,\psi,A] \ = \ i\int\d^4 x 
\bar\psi\left(\p+\frac{1-\gamma_5}{2}\A\right)\psi, \& \& j_a^\mu \& =
i\bar\psi\frac{1-\gamma_5}{2}\gamma^\mu t^a\psi
\eg
with the notation 
\bm{rclcrclcrcl}
A \& = \& A^a t^a,\& \& \Tr\ \! t^a t^b\&  =\&
-\frac{1}{2}\delta^{ab},\& \&  [t^a,t^b]\ =\ {f^{abc}} t^c\\\di 
\gamma_5 \& = & i\gamma^0\gamma^1\gamma^2\gamma^3, & \& \epsilon^{0123}
\& = \& \epsilon^{123}\ =1.  
\eg
The Gauss law operator $G$ in gauge theories is the generator of
time-independent gauge transformations and thus the generator of
gauge transformations on the Fock space. The appropriate gauge in 
this framework is the Weyl gauge $A^0=0$. However, we can not neglect 
terms dependent on $A^0$ at intermediate steps, 
since $\frac{\delta}{\delta A^0} A^0$ contributes. Commutators of $G$ 
(defined on the Fock space) are given by the equal-time commutators of  
\bm{rclcrcl}
G^a(x) \& = \& -i\int \d x_0 D_i^{ab}(x)\frac{\delta}{\delta
  A_i^b(x)}+j^0_a(x), \& \&  D^{ab}_\mu \& = \&
\partial_\mu\delta^{ab}+f^{acb}A_\mu^c,  
\label{Ggauge}\eg
where the first part of $G$ generates time independent gauge
transformations on the (external) gauge field and the second part acts on
fermions. However, for an anomalous gauge 
theory we have to be careful with this identification. Therefore we
start at $t=-\infty$ with vanishing gauge field $A=0$, where the identification is
justified. The
time evolution of $G(t=-\infty,\xv)$ defines
$G$ at later times. Since we work with an external
gauge field, the time evolution operator $U$ is given by 
\bm{rcl}
U(-\infty,x_0) \& \simeq \& \T^*\exp\left\{i\int\d^3 x
\int_{-\infty}^{x_0}\d t A\cdot j(x)\right\},\\\di 
 U(y_0,x_0)\& = \& U(-\infty,x_0) U(y_0,-\infty),
\label{U}\eg 
where $\T^*$ is the Lorentz covariantized time ordered product. 
$U$ has to be regularized and renormalized 
(e.g. \cite{langmann} and references therein). If the representation
of the gauge group is projective (on the Fock space), $U(x_0,x_0)=\id$ can not be
maintained in general. For our purpose it is 
sufficient to discuss the properties of 
\bm{rcccl}
iW[A] \& = \& \ln\langle 0|\T^*\exp\left\{i\int\d^4 x A\cdot j(x)\right\}|0\rangle
 \& = \& \ln \langle 0| U(-\infty,\infty)|0\rangle \label{W[A]}
\eg 
Integrability (consistency) of $W[A]$ is crucial for the following
calculations. Using a gauge covariant regularization of the current $j$
 \cite{mitra}, the integrability of $W[A]$ shows up in the
 (consistent) anomaly equation 
\bm{rcccl}
\CA^a(x)  &\di   = &\di D^{ac}_\mu(x)  \frac{\delta}{\delta
  A^c_\mu(x)}W[A]\& = \& 
\frac{1}{24\pi^2}\epsilon^{\mu\nu\rho\sigma}\partial_\mu
\Tr\left[t^a\left(A_\nu\partial_\rho A_\sigma+ \frac{1}{2}A_\nu
A_\rho A_\sigma\right)\right]. 
\label{anomaly}\eg 
Eq.~(\ref{anomaly}) also indicates the existence of a non-trivial
renormalization factor in \eq{U}. In the following we will only use
the definition of $W[A]$
with a regularized current in \eq{W[A]} and  the anomaly \eq{anomaly} as an
input to calculate commutators and double commutators. 
                    
 At $t=-\infty$ we deal with a free theory. $G(-\infty,\xv)$
(see \eq{Ggauge}) is given by 
\bg 
G(-\infty,\xv) \& = \& G_A(-\infty,\xv)+G_\psi(-\infty,\xv),
\eg
where the fermionic part is the zero component of the consistent
current $J$ 
\bm{rclcrcl}
G_\psi(-\infty,\xv) \& = \& J^0(-\infty,\xv) \& \mbox{with} \& J(-\infty,\xv) \& =
\& \frac{\delta}{\delta A(-\infty,\xv)}\int A\cdot j.
\eg 
Since $ j$ in \eq{W[A]} is gauge field dependent due to the
regularization, $J$ differs from $ j$ for arbitrary time by a term proportional to
$\int  A_\mu\frac{\delta}{\delta A} j^\mu$. $G_A$ is the generator of
time independent gauge transformations on the gauge field and is given
by 
\bg
G_A(-\infty,\xv) \& = \& -i \int\d t D^{ab}_i(t,\xv)
\frac{\delta}{\delta A^b_i(t,\xv)}.
\eg
The equal-time commutators of $G_{A/\psi}(-\infty,\xv)$ are the canonical
commutators, since the gauge field vanishes ($A(-\infty,\xv) =0$). 
Now we define the Gauss law
operator $G(x)$ for arbitrary times $x_0$ as the time evolution of 
$G(-\infty,\xv)$. The fermionic part $G_\psi(x)$ of $G$ is given
by the zero component of the consistent current $J$. 
\bm{rclcrcl}
G_\psi(x) \& = \& J^0(x)  &\mbox{with}\& 
J(x) \& = \& U(x_0,-\infty)J(-\infty,\xv) U(-\infty,x_0).
\label{ferm}\eg
The time evolution of $G_A$ is given by 
\bm{rcccl}
G_A(x) \& = \&
U(x_0,-\infty)G_A(-\infty,\xv)U(-\infty,x_0) \& = \&   -i \int\d t
D^{ab}_i(t,\xv)\del_b^i(t,x),
\label{afield}\eg 
where $\del$ is given by 
\bm{rcl}
\del_b^i(t,x) \& = \& U(x_0,-\infty)U(-\infty,x_0)\frac{\delta}{\delta
  A^b_i(t,\xv)}+\Gamma^i_b(t,x) \\\di  
\mbox{with}\ \ \  \Gamma^i_b(t,x)\& = \& i\theta(x_0-t) J_b^i(t,\xv).
\eg 
$\Gamma^i$ can be seen as a non-trivial
connection for the derivative $\frac{\delta}{\delta A}$ as 
in the two dimensional theory \cite{semenoff}. However, in contrast to
\cite{semenoff} it is not possible to construct $\Gamma^i$ explicitly
in four dimensions.  We relate 
expectation values of $G_A$ to derivatives of $W[A]$ with
respect to the gauge field. The expectation value of 
$G_A$ is connected to $W[A]$ via 
\bm{rcccl}
\llangle G_A(x)\rrangle \& = \& \int \d t\ \theta(x_0-t)
D^{ab}_i(t,\xv)\frac{\delta}{\delta A^b_i(t,\xv)} W[A],
\label{expga}\eg
where the expectation value $\langle\cdots \rangle $ refers to the 
background defined by $W[A]$.   
                 
 In the following we will present equal-time commutators and double
commutators as derivatives  of $W[A]$ with respect to the
gauge field. The components $G_{A/\psi}$ of $G$ are connected to the
following $A$-derivatives
\bm{rcl}
\tilde G_A^a(x) \& = \& \int \d t\ \theta(x_0-t)
D^{ab}_i(t,\xv)\frac{\delta}{\delta A_i^b(t,\xv)},\\\di 
\tilde G_\psi^a(x) \& = \& \int \d t\ \theta(x_0-t)
\partial_0\frac{\delta}{\delta A_0^a(t,\xv)} \ = \  \frac{\delta}{\delta
  A_0^a(x)}.
\label{gtilde}\eg
With eqs. (\ref{expga}),(\ref{gtilde}) we 
derive the well-known relation \cite{fuji} between the time derivative of 
$G=G_A+G_\psi$ and the anomaly (using $A^0=0$)
\bm{rcccccl}
 \partial_{x_0}\llangle G^a(x_0,\xv)\rrangle \& = \&
\partial_{x_0}\tilde G(x) W[A]\& = \& D^{ab}_\mu\frac{\delta}{\delta
  A^b_\mu} W[A] \& = \& \CA^a\label{fuji}.
\eg 
The equal-time commutator of $G_A(x_0,\xv)$ with an operator
\bm{rclcrcl}
\CO(y)\& =\& U(y_0,-\infty)\CO(-\infty,\yv)U(-\infty,y_0) \&
\mbox{with}\& [A,\CO] \& = \& 0
\eg
is given by 
\bg
\Bigl[U(x_0,-\infty)G_A(-\infty,\xv)U(-\infty,x_0),U(x_0,-\infty)\CO(-\infty,\yv)
U(-\infty,x_0)\Bigr].
\eg 
With $D^i\frac{\delta}{\delta A^i} U(x_0,x_0)=0$
we would conclude  
\bg
\Bigl[G_A(x_0,\xv),\CO(x_0,\yv)\Bigr] \& = \& U(x_0,-\infty)
\Bigl[G_A(-\infty,\yv),\CO(-\infty,\yv)\Bigr] U(-\infty,x_0). \label{free}
\eg 
Since we expect a projective representation of the gauge group, $
D^i\frac{\delta}{\delta A^i} U(x_0,x_0)=0$ can not
be assumed. Indeed we have $\tilde G_A(-\infty,\xv)U(x_0,x_0)\neq 0$. 
This can be taken into account by carefully calculating the equal-time
limit    
\bm{rcl}
\& \& \lim_{t\rightarrow x_0} 
G_A(-\infty,\xv)U(-\infty,t)U(x_0,-\infty)\ = \lim_{t\rightarrow x_0} 
G_A(-\infty,\xv)U(x_0,t)\\\di 
& = \&   -i\lim_{p_0\rightarrow\infty}\int \d t e^{i p_0 (t-x_0}
D^{ab}_i(t,\xv)\frac{\delta}{\delta
  A^b_i(t,\xv)}U(-\infty,\infty)  
\label{du}\eg 
The limit $p_0\rightarrow\infty$ in the last line projects on the
terms with $t=x_0$ in $D\frac{\delta}{\delta A}U(-\infty,\infty)$. Thus we conclude 
\bg
\llangle\Bigl[G_A(x),\CO(y)\Bigr]_{ET}\rrangle \& = \& -i\lim_{p_0
 \rightarrow \infty}
\int \d x_0 e^{ip_0 (x_0-y_0)} D^{ab}_i(x)\frac{\delta}{\delta
  A^b_i(x)} \llangle \CO(y)\rrangle.
\label{comofga}\eg
For operators $\CO$ containing $\frac{\delta}{\delta A}$ we have
additional terms proportional to $\frac{\delta}{\delta A}[G_A]$.  
Eq.~(\ref{comofga}) has the form of a BJL-limit \cite{bjl}, which connects the 
time-ordered product of two (bosonic) operators $A,B$ with their equal
time commutator. We have formally 
\bg
\int\d x_0 e^{ip_0 x_0}\partial_{x_0}\T A(x) B(0) \& = \&
\int\d x_0 e^{ip_0 x_0}\partial_{x_0}\left[
\theta(x_0)A(x)B(0)+\theta(-x_0)B(0)A(x)\right]\\\di  
\& = \& [A(x),B(0)]_{x_0=0}+\int\d x_0 e^{ip_0 x_0}\T
\partial_{x_0}A(x)B(0).
\label{bjl1}\eg 
Providing a suitable regularization for the operators $A$ and $B$, the
second term vanishes in the limit $p_0\rightarrow\infty$. The
extension to double commutators is obvious \cite{rothe2}. 
\bg
\Bigl[ A(x),[B(y),C(0)]\Bigr]_{ET} 
\& = \& \lim_{p_0\rightarrow\infty}\lim_{q_0\rightarrow\infty}
\int\d x_0\d y_0 e^{ip_0 x_0} e^{iq_0 y_0}\partial_{x_0}\partial_{y_0}
\T A(x) B(y)C(0) 
\label{bjl2}\eg
where the subscript $ET$ denotes equal-time. In \eq{bjl2} 
we have to perform first the $q_0\rightarrow\infty$
limit. Moreover this only provides a proper definition of the
double commutator, if 
the regularization of the time ordered product $\T A(x)
B(y)C(0)$ does not affect the $\theta$-functions. However,
performing the BJL-limit perturbatively in Feynman integrals, this
condition is violated by the exchange of integration and limit
procedure. The diagrammatic calculation of double commutator with 
formally equivalent BJL-limits gives not the same result in
general (e.g. fermionic currents \cite{rothe},\cite{kubo},\cite{rothe2}), which
indicates the failure of the iterative BJL-limit. This is the reason for
the violation of the Jacobi identity within the iterative
BJL-procedure \cite{kubo},\cite{double1}. 
                            
 Commutators containing $G_A$ can be
expressed as the limit of time-ordered products with the BJL-method.
With use of eqs. (\ref{afield}),(\ref{comofga}) we get ($\CO$ does not
contain $\frac{\delta}{\delta A}$)
\bg 
\llangle\Bigl[G_A(x),\CO(0)\Bigr]\rrangle \& = \& 
\lim_{p_0\rightarrow\infty}
\int\d x_0 e^{ip_0 x_0}\partial_{x_0}\llangle \T^* G_A(x)
\CO(0)\rrangle\\\di 
\& = \& -i\lim_{p_0 \rightarrow \infty}\d t e^{ip_0 t} D^{ab}_i(t,\xv)\frac{\delta}{\delta
  A^b_i(t,\xv)} \langle \CO(0)\rangle,
\label{comga}\eg 
The covariantized time-ordering $T^*$ appears naturally in the
definition of the BJL-limit, if $G_A$ is involved. However, if  
\bg
\langle 0| \T^* \CO(y)e^{i\int A\cdot j}|0\rangle \& = \& 
\langle 0| \T \CO(y)e^{i\int A\cdot j}|0\rangle, 
\label{t-t}\eg
the results do not depend on the use of the usual time-ordering $\T$ or
$\T^*$. The operators $\CO$ mentioned here have the property
\eq{t-t}. 
               
 Eq.~(\ref{comga}) can be extended 
to arbitrary commutators and double commutators of
$G_A$, if we take into account derivatives of $G_A$ with respect to
the gauge field.  
                     
 We want to emphasize that \eq{afield} and the 
BJL-limit \eq{comga} coincide with the BJL formulae for $-D\cdot E$ in
a chiral theory with quantized gauge field, where only fermionic loops
are taken into account \cite{jo2}. 
             
\mysection{The Algebra of Components of $G$}
In the following we use the properties of $G_{A/\psi},W[A]$ and
the BJL-method for the calculation of the equal-time
commutators of $G_{A/\psi}$. The results are derived only from the
consistent anomaly. It is well known, that the Schwinger terms of 
the different commutators are related by functional
derivatives of the anomaly (e.g. \cite{tsu}). Given these relations we only have to
calculate one Schwinger
term from the (consistent) anomaly $\CA$. First we derive the 
relations between Schwinger terms within the formalism introduced in
section~\ref{gauss}. It follows by \eq{comga}
\bg
\llangle\left[G^a(x),G_\psi^b(y)\right]_{ET}\rrangle &\di    = &\di    
  -i\lim_{p_0\rightarrow \infty}\int \d x_0 e^{ip_0 (x_0-y_0)}
\partial^x_0\tilde G^a(x)\llangle J^0_b(y)\rrangle\\\di  
\& = \& -i\lim_{p_0\rightarrow \infty}\int \d x_0 e^{ip_0 (x_0-y_0)}
\frac{\delta}{\delta A^b_0(y)} 
\partial_0^x \tilde G_\psi^a(x)W[A],
\eg
where the integrability of $W[A]$ was used by commuting the
derivatives with respect to $A$. With \eq{fuji} we conclude 
\bg
\llangle\left[G^a(x),G_\psi^b(y)\right]_{ET}\rrangle 
&\di    = &\di     -i\lim_{p_0\rightarrow \infty}\int \d x_0 e^{ip_0 (x_0-y_0)}
\frac{\delta}{\delta A^b_0(y)} \CA^a(x)\\\di  
&\di      &\di    +i\lim_{p_0\rightarrow \infty}\int \d x_0 e^{ip_0 (x_0-y_0)}
\frac{\delta}{\delta A^b_0(y)} 
f^{acd}A_0^c(x)\frac{\delta}{\delta A_0^d(x)} W[A]\\\di  
 &\di    = &\di    \llangle iG^{[a,b]}(x)\rrangle\delta(\xv-\yv)-
i\int \d x_0 \frac{\delta}{\delta A^b_0(y)}
\CA^a(x). 
\label{relation1}\eg 
The relations between the other commutators follow similarly. We quote
the results 
\bg
\llangle\left[G^a(x),G_A^b(y)\right]_{ET}\rrangle &\di    = &\di    
i\llangle G_A^{[a,b]}(x)\rrangle\delta(\xv-\yv)
+i\int \d x_0 D^{bd}_j(y)\frac{\delta}{\delta 
\partial_0 A_j^d(y)}\CA^a(x)\\\di  

\llangle\left[G^a(x), G_\psi^b(y)\right]_{ET}\rrangle &\di    = &\di  
i\llangle G_\psi^{[a,b]}(x)\rrangle\delta(\xv-\yv)
-i\int \d x_0\frac{\delta}{\delta A_0^b(y)}\CA^a(x)\\\di 

\llangle \left[G^a(x),G^b(y)\right]_{ET}\rrangle &\di    = &\di    
i\llangle G^{[a,b]}(x)\rrangle\delta(\xv
-\yv)\\\di 
\&  \&   -i\int \d x_0 \left[\frac{\delta}{\delta A_0^b(y)}
-D^{bc}_i(y)\frac{\delta}{\delta \partial_0 A_i^c(y)}\right]\CA^a(x).
\label{result}\eg 

Now we calculate the commutator $[G_A^a,G_A^b]$ with \eq{result} and
considerations concerning symmetry properties. The 
anomalous Schwinger terms are connected to terms in $W[A]$ containing
at least cubic powers of the gauge field. Hence the Schwinger terms 
contain at least linear powers of the gauge field. The only term with
the correct symmetry properties is 
\bg 
\llangle\left[G_A^a(x),G_A^b(y)\right]_{ST}\rrangle \& = \&
q\epsilon^{ijk}D^{ac}_i(x)D^{bd}_j(y)\Tr\left[\{t^c,t^d\}A_k\right]\delta(\xv-\yv),
\label{gaga}\eg 
which is connected to a $U(1)$-curvature in field space \cite{jo},\cite{banerjee},
\cite{semenoff}. In the present approach it follows by the
observation, that the Schwinger term \eq{gaga} is directly related to  
$D^i\frac{\delta}{\delta A^i}U(x_0,x_0)\neq 0$, where 
$U(x_0,x_0)$ defines a loop in the gauge group. 
It remains to determine the constant $q$. Eq.~(\ref{result})
establishes the relation between the Schwinger term \eq{gaga} and
$[G_\psi^a,G_A^a]_{ET}$
\bg
\llangle\left[G_A^a(x),G_A^b(y)\right]_{ST}\rrangle &\di    = &\di   
-\llangle\left[G_\psi^a(x),G_A^b(y)\right]_{ET}\rrangle  \\\di 
\&  \& 
-\frac{i}{24\pi^2}\epsilon^{ijk}D_j^{bd}(y)\Tr\left[\{t^c,t^d\}\partial_i
A_k(x)\right]\delta(x-y). 
\label{relation}\eg 
It follows from eqs. (\ref{gaga}),(\ref{relation}) that 
\bg
\llangle\left[G_\psi^a(x),G_A^b(y)\right]_{ET}\rrangle \& \sim \& -q
 D^{bd}_j(y)\Tr\left[\{t^a,t^d\}A_k(x)\right]\partial_i^x\delta(\xv-\yv).
\label{structure}\eg 
Thus we only have to evaluate these terms contributing to
$[G_\psi^a,G_A^a]_{ST}$ to determine $q$. For this purpose 
we introduce the covariant current $\tilde J$, which differs from the
consistent current by the Bardeen-Zumino polynomial $\Delta J$ 
\cite{bardeen}. 
\bm{rclcrcl}
\llangle J^\mu_a\rrangle \&= \&  \llangle \tilde J^\mu_a\rrangle
-\Delta J^\mu_a \&  \mbox{with}\&  \Delta J^\mu_a \& = \& 
\frac{1}{24\pi^2}\epsilon^{\mu\nu\rho\sigma}\Tr\left[t^a\{A_\nu,\partial_\rho
A_\sigma\}+\frac{3}{2}A_\nu A_\rho A_\sigma\right].  
\label{covariant}\eg 
We use the gauge covariance of $\langle \tilde J\rangle$ in the
following derivation. With \eq{covariant} we have 
\bg
\llangle\left[G_\psi^a(x),G_A^b(y)\right]_{ET}\rrangle  \& = \& i
\lim_{p_0\rightarrow\infty}\int\d y_0 e^{i p_0 (y_0-x_0)}
D^{bd}_j(y)\frac{\delta}{\delta A_j^d(y)}\llangle J^0_a(x)\rrangle\\\di
\&  = \&  i
\lim_{p_0\rightarrow\infty}\int\d y_0 e^{i p_0 (y_0-x_0)}
D^{bd}_j(y)\frac{\delta}{\delta A_j^d(y)}\left[\llangle \tilde
  J^0_a(x)\rrangle-\Delta J^0_a(x)\right].
\eg
It follows from the covariance of $\langle\tilde J\rangle$ that the
first term on the right hand side does not contribute to
\eq{structure}. 
Thus, only taking into account
terms which can contribute to \eq{structure}, we get 
$(\epsilon^{ijk}=\epsilon^{0ijk})$ 
\bg
\llangle\left[G_\psi^a(x),G_A^b(y)\right]_{ET}\rrangle  \& \sim \&  -i
\lim_{p_0\rightarrow\infty}\int\d y_0 e^{i p_0 (y_0-x_0)}
D^{bd}_j(y)\frac{\delta}{\delta A_j^d(y)}\Delta J^0_a(x)\\\di 
\& = \& 
-\frac{i}{24\pi^2}\epsilon^{ijk}D^{bd}_j\Tr\left[\{t^c,t^d\}A_k(x)\right]
\partial_i^x\delta(\xv-\yv).
\eg
This determines $q=\frac{i}{24\pi^2}$ and we have finally 
\bg 
\llangle \left[G_A^a(x),G_A^b(y)\right]_{ET}\rrangle \& = \& i\llangle
G_A^{[a,b]}(x)\rrangle
\delta(\xv-\yv)\\\di 
\&  \& +\frac{i}{24 \pi^2}\epsilon^{ijk}D^{ac}_i(x)
D^{bd}_j(y)\Tr\Bigl[\{t^c,t^d\}A_k\Bigr]\delta(\xv-\yv).
\label{schwinger}\eg  
Together with eq.~(\ref{result}) this determines all commutators. 
The results coincide with the literature (e.g. \cite{jo},\cite{tsu}). 

\mysection{Double Commutators}
In the derivation of \eq{schwinger} we used relations
only valid as expectation values (see
eqs.~(\ref{relation})-(\ref{schwinger})). Thus we expect,
that it is not possible to calculate the double commutator 
$\Bigl[G_A^a,[G_A^b,G_A^c]\Bigr]$ iteratively. 
Using the form of $G_A$ (see \eq{afield}) we conclude
\bg
&\di     &\di    \llangle\left[G_A^a(x),\left[G_A^b(y),
G_A^c(z)\right]\right]_{ET}\rrangle\\\di   
&\di    = &\di    
 \llangle\left[G_A^a(x),iG_A^{[b,c]}(y)\right]_{ET}\rrangle
\delta(\yv-\zv)\\\di 
\&  \& +i\int_{t,t',t''}\Biggl\{\left[D^{ad}_i(t,\xv)\frac{\delta}{\delta A_i^d(t,\xv)} 
\left(D^{be}_j(t',\yv)D^{cf}_k(t'',\zv)\right)\right]
\llangle\left[ \del_e^j(t',y),\del_f^k(t'',z) \right] \rrangle\\\di  
&\di       &\di    +D^{ad}_i(t,x)D^{be}_j(t',y)D^{cf}_k(t'',z)
\llangle\Bigl[\del_d^i(t,x),\left[\del_e^j(t',y),\del_f^k(t'',z)\right]\Bigr]\rrangle
\Biggr\}.
\label{nonit}\eg
The first two terms follow easily with eq.~(\ref{schwinger}), since
they only involve derivatives of 
$\tilde G_A$ with respect to the gauge field. 
However, the last term can not be calculated 
with the known commutators. It
follows from algebraic considerations, that it has to vanish, if 
the Jacobi identity is fulfilled. The
structure of the double commutator
\bg
K^{ijk}_{def}(\xv,\yv,\zv) &\di \!   = \& \! i\int_{t,t',t''}
D^{ad}_i(t,x)D^{be}_j(t',y)D^{cf}_k(t'',z)
\llangle\Bigl[\del_d^i(t,x),\left[\del_e^j(t',y),\del_f^k(t'',z)\right]\Bigr]\rrangle
\label{delta^3}\eg
follows by dimensional analysis as 
\bg
K^{ijk}_{def}(\xv,\yv,\zv) &\di    = &\di    q\epsilon^{ijk}D^{ad}_i(x)D^{be}_j(y)
D^{cf}_k(z)\Tr\left[t^d\{t^e,t^f\}\right]
\delta(\xv-\yv)\delta(\yv-\zv).
\eg
The iterative result obtained with \eq{schwinger} is
$q=\frac{1}{24\pi^2}$. The Jacobi identity is only fulfilled when 
$q=0$. Performing the BJL-limit we get for the last 
term in \eq{nonit} 
\bg 
K_{def}^{ijk}(\xv,\yv,\zv) 
\& = \& 
-\lim_{p_0\rightarrow \infty}\lim_{q_0\rightarrow\infty}
\int \d x_0 \d y_0\d z_0 e^{ip_0 x_0}e^{iq_0 y_0} \theta(-z_0)
D^{ad}_i(x)D^{be}_j(y)D_k^{cf}(z)\\\di 
\& \& \cdot\left[\frac{\delta}{\delta A^d_i(x)} 
\frac{\delta}{\delta A^e_j(y)}\frac{\delta}{\delta A^f_k(z)}W[A]\right]\\\di 
 \& = \& 0.
\label{double1}\eg 
The only terms of 
\bgo 
\frac{\delta}{\delta A^d_i(x)} 
\frac{\delta}{\delta A^e_j(y)}\frac{\delta}{\delta A^f_k(z)}W[A],
\ego
which contribute to \eq{double1} are proportional to
$\partial_0\delta(z_0-y_0)\delta(z_0-x_0), 
\partial_0\delta(y_0-z_0)\delta(z_0-y_0)$ and 
$\partial_0\delta(z_0-y_0)\delta(x_0-x_0)$. The group structure is
similar to \eq{schwinger}. It follows with the integrability of 
$W[A]$ that \eq{double1} is proportional to 
\bg
\Tr \left[ t^d\{t^e,t^f\}\right]\Bigl(\partial_0\delta(z_0-y_0)\delta(z_0-x_0)+
\mbox{cycl. perms. of } (x_0,y_0,z_0)\Bigr) \& = \& 0.
\label{0}\eg 
In an iterative BJL-limit one would only take into account one of the terms
proportional to $\partial_0\delta(z_0-y_0)\delta(z_0-x_0)$ and $
\partial_0\delta(x_0-z_0)\delta(x_0-y_0)$. However, only the sum of
these two terms add up to zero in the limit, which is the reason for the 
violation of the Jacobi identity within an iterative calculation (the
term proportional to $\partial_0\delta(x_0-z_0)\delta(y_0-z_0)$
is suppressed with $p_0/(p_0+q_0)$ in the BJL-limit in \eq{double1}). 
With \eq{double1} we have finally 
\bg
\&  \& \llangle\Bigl[G_A^a(x),\left[G_A^b(y),G_A^c(z)\right]
\Bigr]_{ET}\rrangle \\\di 
&\di    = &\di    \llangle\left[G_A^a(x),iG_A^{[b,c]}(y)
\right]_{ET}\rrangle\delta(\yv-\zv)\\\di 
\&  \& +\frac{1}{24\pi^2}\epsilon^{jkl}\int\d
t\ D_i^{ad}(x)\left[\frac{\delta}{\delta A^d_i(x)}
\left(D^{be}_j(y)D^{cf}_k(z)\right)\right]\Tr[\{t^e,t^f\}A_l]\delta(\yv-\zv)\\\di  
&\di    = &\di    \llangle\Bigl[G_A^a(x),[G_A^b(y),
G_A^c(z)]\Bigr]_{it}\rrangle\\\di 
\&  \& -\frac{1}{24\pi^2}\epsilon^{ijk}D^{ad}_i(x)D^{be}_j(y)
D^{cf}_k(z)\Tr[t^d\{t^e,t^f\}]\delta(\xv-\yv)\delta(\yv-\zv)
\end{array}\label{nichtit}\end{eqnarray}
The double commutator with the subscript $it$ is the iterative double
commutator. 
                
 Now we proceed as in the case of the commutators. We use the
anomaly and the double commutator eq.~(\ref{nichtit}) to
calculate the other double commutators contributing to the algebra of
$G$. As an important first step we prove, that double commutators 
with the structure $\Bigl[G^a_{A/\psi},[G^b_{A/\psi},G^c]\Bigr]$ agree with the
iterative results. We derive with use of the notation \eq{gtilde} 
\bg
& \& \left \langle\left[G_{A/\psi}^a(x),\left[G_\psi^b(y),G^c(0)\right]
\right]_{ET}\rrangle
\\\di 
 \& = \&
-\lim_{p_0\rightarrow \infty}\lim_{q_0\rightarrow
   \infty}p_0 q_0 \int \d y_0\d x_0 e^{ip_0 x_0}e^{iq_0 y_0} 
 \llangle\T^* G_{A/\psi}^a(x) G_{\psi}^b(y)G^c(0)\rrangle\\\di 
\& = \& i\lim_{p_0\rightarrow \infty}p_0\int\d x_0 e^{ip_0 x_0}
\tilde G_{A/\psi}^a(x)\lim_{q_0\rightarrow
   \infty} q_0 \int \d y_0e^{iq_0 y_0} \llangle\T^*
 G_\psi^b(y)G^c(0)\rrangle\\\di 
\& = \& \lim_{p_0\rightarrow \infty}p_0\int\d x_0 e^{ip_0 x_0}
\tilde G_{A/\psi}^a(x)\lim_{q_0\rightarrow \infty} q_0 \int \d
y_0e^{iq_0 y_0} \tilde G_\psi^b(y)\tilde G^c(0) W[A].
\eg
Here we used, that derivatives of $G^a_{A/\psi}$ with respect to the gauge field 
vanish in the BJL-limit with at least $p_0/(p_0+q_0)\rightarrow 0$.
Performing the limit $q_0\rightarrow\infty$ we get 
\bg
& \& \llangle\left[G_{A/\psi}^a(x),\left[G_{\psi}^b(y),G^c(0)\right]
\right]_{ET}\rrangle
\\\di 
\& = \& i\lim_{p_0\rightarrow\infty}\int\d x_0 e^{i p_0
   x_0}\partial^x_0\tilde G^a_{A/\psi}(x)\llangle G_{\psi}^{[b,c]}(0)\rrangle\delta(\yv)\\\di 
\&  \& -\lim_{p_0\rightarrow \infty}p_0\int\d x_0 e^{ip_0 x_0} \tilde G_{A/\psi}^a(x)
\left[\lim_{q_0\rightarrow
   \infty}q_0 \int \d y_0 e^{iq_0 y_0}
 \tilde G_{\psi}^b(y)\int \d t \theta(-t)\CA^c(t,0)\right]\\\di 
\& = \& \llangle
[G^a_{A/\psi}(x),iG_{\psi}^{[b,c]}(0)]_{ET}\rrangle\delta(\yv)+
\int\d x_0\partial_0\tilde G_{A/\psi}^a(x)
\int \d y_0 \frac{\delta}{\delta A^b_0(y)}\CA^c(0).
\label{it}\eg
For $G^a_{A/\psi}=G^a_\psi$ the second term in the last line vanishes.
For $G^a_{A/\psi}=G^a_A$ we have 
\bgo
\int \d x_0\partial_0\tilde G^a_A(x) \& = \& \int\d x_0
D_i^{ad}(x)\frac{\delta}{\delta A_i^d(x)}.
\ego
Thus \eq{it} is the iterative result. In the derivation we used 
\bg
\tilde G^c(x)W[A] \& = \& \int \d t\ \theta(x_0-t)\left(\CA^c(t,\xv)-f^{cde}
  A^d_0(t,\xv)\frac{\delta}{\delta A_0^e(t,\xv)}W[A]\right).
\label{gotoit}\eg 
The Schwinger term in \eq{it} is simply given by functional
derivatives of the anomaly in contrast to the Schwinger term of
$\Bigl[G^a_A,[G^b_A,G^c_A]\Bigr]$. It is directly related to the 
2-cocycle in the algebra of the Gauss
law operator $G$. 
                  
 Applying the derivation of \eq{it} to the double commutator
$\Bigl[G^a_{A/\psi},[G^b_{A},G^c]\Bigr]$, we get 
\bg
& \& \left \langle\left[G_{A/\psi}^a(x),\left[G_A^b(y),G^c(0)\right]
\right]_{ET}\rrangle
\\\di 
 \& = \&
-\lim_{p_0\rightarrow \infty}\lim_{q_0\rightarrow
   \infty}p_0 q_0 \int \d y_0\d x_0 e^{ip_0 x_0}e^{iq_0 y_0} 
 \llangle\T^* G_{A/\psi}^a(x)G_{A}^b(y)G^c(0)\rrangle\\\di 
\& = \& i\lim_{p_0\rightarrow \infty}p_0\int\d x_0 e^{ip_0 x_0}
\tilde G_{A/\psi}^a(x)\lim_{q_0\rightarrow
   \infty} q_0 \int \d y_0e^{iq_0 y_0} \llangle\T^*
 G_A^b(y)G^c(0)\rrangle\\\di 
\&  \& -\int \d t \theta(-t)D^{[b,c]d}_i(t,0)
\left[\frac{\delta}{\delta A_i^d(t,0)}\tilde
  G_{A/\psi}^a(x)\right]W[A]\delta(\yv)\\\di 
\& = \& \llangle
[G_{A/\psi}(x),iG_{A}^{[b,c]}(0)]_{ET}\rrangle\delta(\yv)-
\int\d x_0 \partial_0\tilde G_{A/\psi}^a(x)\int \d y_0 D_j^{bd}(y)
\frac{\delta}{\delta\partial_0 A^d_j(y)}\CA^c(0).
\label{it1}\eg 
It follows with eqs.~(\ref{it}),(\ref{it1}), that double
commutators of the form $\Bigl[G^a_{A/\psi},[G^b_{A/\psi},G^c]\Bigr]$ agree with the
iterative results. Therefore they can be calculated from the known 
commutators eqs.~(\ref{result}),(\ref{schwinger}). The algebra for
arbitrary combinations of $G_{A/\psi}$ is determined with the double
commutators eqs.~(\ref{nichtit}),(\ref{it}),(\ref{it1}) and the result
for $\Bigr[G^a_\psi,[G^b_{A},G^c_{A}]\Bigr]$. We calculate 
\bg
\llangle\left[G_\psi^a(x),\left[G_A^b(y),G_A^c(0)\right]\right]_{ET}\rrangle 
\!\! &\di    =  &\di \!\!   
\llangle\left[G_\psi^a(x),iG_A^{[b,c]}(0)\right]_{ET}\rrangle\delta(\yv)\\\di  
&\di     &\di    -\lim_{p_0\rightarrow \infty}\lim_{q_0\rightarrow \infty} 
\int \d x_0\d y_0\d z_0 e^{ip_0 x_0}e^{iq_0 y_0}\theta(-z_0)\\\di  
&\di      &\di    \cdot 
D^{bd}_i(y)D^{ce}_j(z)\frac{\delta}{\delta A_i^d(y)}\frac{\delta}{\delta A_j^e(z)}
\partial_0\frac{\delta}{\delta A_0^a(x)}W[A].
\eg
Using  
\bgo 
\partial_0\frac{\delta}{\delta A^0_a(x)}W[A] \& = \&
\CA^a(x)-\left[D^{af}_k(x)
\frac{\delta}{\delta A_k^f(x)}+f^{agf}A_0^g(x)\frac{\delta}{\delta
  A_0^f(x)}\right]W[A]
\ego
it follows 
\bg
\llangle\left[G_\psi^a(x),\left[G_A^b(y),G_A^c(0)\right]\right]_{ET}\rrangle
\!\!&\di    = &\di \!\!   
\llangle\left[G_\psi^a(x),iG_A^{[b,c]}(0)\right]_{ET}\rrangle\delta(\yv)\\\di  
&\di    &\di  -\int\d x_0\d y_0
D^{bd}_i(y)D^{ce}_j(z)\frac{\delta}{\delta \partial_0 A_i^d(y)}
\frac{\delta}{\delta A_j^e(z)}\CA^a(x)\\\di 
&\di    &\di  +\int\d x_0\d y_0 D^{bd}_i(y)D^{ce}_j(z)
\frac{\delta}{\delta A_i^d(y)}\frac{\delta}
{\delta \partial_0 A_j^e(z)}\CA^a(x)\\\di 
&\di      &\di    +\lim_{p_0\rightarrow \infty}\lim_{q_0\rightarrow \infty}
\int \d x_0\d y_0 d z_0 e^{ip_0 x_0}e^{iq_0 y_0}\theta(-z_0)\\\di  
&\di     &\di    \cdot D^{bd}_i(y)D^{ce}_j(z)\frac{\delta}{\delta A_i^d(y)}
\frac{\delta}{\delta A_j^e(z)}
D^{af}_k(x)\frac{\delta}{\delta A_k^f(x)}W[A].\label{zwischen}
\eg
The terms containing functional derivatives of the anomaly have the
form 
\bg
\&  \& \int\d x_0\int\d y_0 D^{bd}_i(y)D^{ce}_j(z)\left[
\frac{\delta}{\delta A_i^d(y)}\frac{\delta}{\delta \partial_0 A_j^e(z)}
-\frac{\delta}{\delta \partial_0 A_i^d(y)}\frac{\delta}{\delta A_j^e(z)}\right]
\CA^a(x)\\\di 
 \& = \& \frac{1}{24 \pi^2}
\epsilon^{ijk}D^{bd}_i(y)D^{ce}_j(z)\partial^x_k 
\Tr \left[t^a\{t^d,t^e\}\right]\delta(\xv-\yv)\delta(\xv-\zv).
\eg
The last term on the right hand side of \eq{zwischen} consists of
equal-time commutators of
the form $[\del_{[a,g]},\del_h]$ (see \eq{schwinger}) and terms
proportional to $\delta^3_A W[A]$, which vanish in the BJL-limit. 
Performing the BJL-limit, we are led to 
\bg
 &\di     &\di
 \llangle\left[G_\psi^a(x),\left[G_A^b(y),G_A^c(z)\right]\right]_{ET}
\rrangle\\\di  
&\di    = &\di    \llangle\left[G_\psi^a(x),iG_A^{[b,c]}(y)\right]_{ET}\rrangle
\delta(\yv-\zv) +\frac{1}{24 \pi^2}
\epsilon^{ijk}D^{bd}_i(y)D^{ce}_j(z)\partial^x_k \\\di 
\& \& \cdot\Tr \left[t^a\{t^d,t^e\}\right]\delta(\xv-\yv)\delta(\xv-\zv)-i\int\d t \d t' 
D^{bd}_i(t,\yv)D^{ce}_j(t',\zv)\\\di 
\&  \& \cdot\Bigl\{
\llangle\left[\del^i_{[a,d]}(t,y),\del^j_e(t',z)\right]\rrangle
\delta(\xv-\yv)-\llangle\left[\del^j_{[a,e]}(t',z),
\del^i_d(t,y)\right]\rrangle\delta(\xv-\zv)\Bigr\}\\\di 

&\di    = &\di    \llangle\left[G_\psi^a(x),iG_A^{[b,c]}(y)
\right]_{ET}\rrangle\delta(\yv-\zv)\\\di  
&\di    &\di   +\frac{1}{24 \pi^2}
\epsilon^{ijk}D^{bd}_i(y)D^{ce}_j(z)\partial^x_k 
\Tr \left[t^a\{t^d,t^e\}\right]\delta(\xv-\yv)\delta(\xv-\zv)\\\di 
&\di     &\di    -\epsilon^{ijk}\frac{1}{24\pi^2}D^{bd}_i(y)D^{ce}_j(z)
\Tr\Bigl[\Bigl(\{[t^a,t^d],t^e\}+\{[t^a,t^e],t^d\}\Bigr)A_k(x)\Bigr]
\delta(\xv-\yv)\delta(\xv-\zv)\\\di 
\& = \&
\llangle\left[G_\psi^a(x),iG_A^{[b,c]}(y)\right]_{ET}\right
\rangle\delta(\yv-\zv)\\\di  
&\di    &\di   +\frac{1}{24 \pi^2}
\epsilon^{ijk}D^{bd}_i(y)D^{ce}_j(z) D^{af}_k(x) 
\Tr \left[t^f\{t^d,t^e\}\right]\delta(\xv-\yv)\delta(\xv-\zv).
\end{array}\label{nichtit5}\end{eqnarray}
The non-iterative part in \eq{nichtit5} is exactly the same as for the
double commutator \eq{nichtit} with a relative minus sign. We conclude
\bg
\llangle\left[G^a(x),\left[G_A^b(y),G_A^c(z)\right]\right]_{ET}\rrangle
\& = \&
\llangle\left[G^a(x),\left[G_A^b(y),G_A^c(z)\right]\right]_{it}\rrangle.
\label{it2}
\eg
We want to
emphasize, that the non-iterative part of eq.~(\ref{nichtit5}) is
also connected to the fact, that the result for the 
commutator $[G^a_A,G^b_A]$ (see \eq{schwinger}) is not
valid on the operator level. With eqs.~(\ref{it}),(\ref{it1}),(\ref{it2}) it
follows, that double
commutator containing at least one full $G$ are given by the
iterative results. We collect the results 
\bg
\llangle\left[G_{A/\psi}^a(x),\left[G^b_{A/\psi}(y),G^c(z)\right]\right]_{ET}\rrangle
\& = \&
\llangle\left[G_{A/\psi}^a(x),\left[G_{A/\psi}^b(y),G^c(z)\right]\right]_{it}\rrangle\\\di

\llangle\left[G^a(x),\left[G^b_{A/\psi}(y),G^c_{A/\psi}(z)\right]\right]_{ET}\rrangle
\& = \&
\llangle\left[G^a(x),\left[G^b_{A/\psi}(y),G^c_{A/\psi}(z)\right]\right]_{it}\rrangle.
\label{it3}
\eg
Eq.~(\ref{it3}), 
the commutators \eq{result},(\ref{schwinger}) and the
double commutator \eq{nichtit} fix all double commutators, as already
mentioned. 
The algebra of two $G_A$ and one $G_\psi$ is given by 
\bg 
\llangle\left[G_A^a(x),\left[G_A^b(y),G_\psi^c(z)\right]\right]_{ET}\rrangle   
\! &\di    = &\di \!   \llangle
\left[G_A^a(x),\left[G_A^b(y),G_\psi^c(z)\right]\right]_{it}\rrangle
+\frac{1}{24\pi^2}
\epsilon^{ijk} D^{ad}_i(x)\\\di   
&\di     &\di \cdot D^{be}_j(y)
D^{cf}_k(z)\Tr\left[t^d\{t^e,t^f\}\right]\delta(\xv-\yv)\delta(\yv-\zv)\\\di 

\llangle\left[G_\psi^a(x),\left[G_A^b(y),G_A^c(z)\right]\right]_{ET}\rrangle  
&\di    = &\di
\llangle\left[G_\psi^a(x),iG_A^{[b,c]}(y)\right]_{it}\rrangle +  
\frac{1}{24\pi^2}\epsilon^{ijk} D^{ad}_i(x)\\\di   
&\di     &\di \cdot D^{be}_j(y)
D^{cf}_k(z)\Tr\left[t^d\{t^e,t^f\}\right]\delta(\xv-\yv)\delta(\yv-\zv)  
\label{ueber3}\eg
The algebra of two $G_\psi$ and one $G_A$ is given by 
\bg
\llangle\left[G_\psi^a(x),\left[G_A^b(y),G_\psi^c(z)\right]\right]_{ET}\rrangle
&\di  =   &\di  -\frac{1}{24\pi^2}\epsilon^{ijk}D^{ad}_i(x)D^{be}_j(y)
D^{cf}_k(z)\\\di 
\&  \& \cdot \Tr\left[t^d\{t^e,t^f\}\right]\delta(\xv-\yv)\delta(\yv-\zv)\\\di 

\llangle\left[G_A^a(x),\left[G_\psi^b(y),G_\psi^c(z)\right]\right]_{ET}\rrangle
&\di  = &\di
\llangle\left[G_A^a(x),\left[G_\psi^b(y),G_\psi^c(z)\right]\right]_{it}\rrangle\\\di  
&\di    &\di -\frac{1}{24\pi^2}\epsilon^{ijk}D^{ad}_i(x)D^{be}_j(y)
D^{cf}_k(z) \Tr\left[t^d\{t^e,t^f\}\right]\\\di 
\& \& \cdot \delta(\xv-\yv)\delta(\yv-\zv) 
\end{array}\label{ueber4}\end{eqnarray}
The algebra of the fermionic currents is given by 
\bg
\llangle\left[G_\psi^a(x),\left[G_\psi^b(y),G_\psi^c(z)\right]\right]_{ET}\rrangle 
&\di   = &\di
\llangle\left[G_\psi^a(x),iG_\psi^{[b,c]}(y)\right]_{ET}\rrangle
\delta(\yv-\zv)
 \\\di  
&\di     &\di  +\frac{1}{24\pi^2}\epsilon^{ijk}D^{ad}_i(x)D^{be}_j(y)
D^{cf}_k(z) \Tr\left[t^d\{t^e,t^f\}\right] \\\di 
\& \& \cdot \delta(\xv-\yv)\delta(\yv-\zv) 
\label{currents}\eg 
It can be shown explicitly in a long, but straightforward calculation,
that the Jacobi identity is fulfilled for all these results. Since we
have related the double commutators to derivatives of $W[A]$ with respect
to the gauge field, this provides a convenient way to prove the Jacobi identity
(integrability of $W[A]$). 
                    
\mysection{Discussion}
In this paper a purely algebraic BJL-type derivation of the algebra 
of the Gauss  law operator $G$ in chiral gauge theories was given. The
theory has been defined in an external gauge field (adiabatic
solution). The expectation values of commutators and double
commutators can be expressed as 
derivatives of $W[A]$ with respect to the gauge field. 
All Schwinger terms follow with the anomaly without any further
assumption. 
             
The Schwinger terms depend on the properties of 
the time evolution operator $U$, which defines the underlying 
theory. If the representation of the gauge group is projective (on the
Fock space), we have 
$D^i\frac{\delta}{\delta A^i} U(x_0,x_0)\neq 0$ 
(see section~\ref{gauss}). This is 
the reason for the Schwinger term in the commutator $[G_A^a,G_A^b]$ 
(\eq{schwinger}). We want to emphasize that the explanation of this
Schwinger term is based purely on an adiabatic solution of the theory in an external
gauge field. There is no need to introduce a quantized gauge field 
to obtain this result. 
                
The commutators (\eq{result} and
\eq{schwinger}) coincide with the BJL-results obtained via the
explicit calculation of 
diagrams \cite{jo}, or path integral methods \cite{tsu}, where only
fermionic loops give contributions. In contrast to these
approaches we do not need to calculate diagrams nor do we need
further assumptions as in \cite{tsu}. The algebraic derivation in
\cite{tsu} was based on the assumption, that $[G_A^a,G_A^b]_{ST}=0$ 
for $G$ defined with the covariant current. Adding the Bardeen-Zumino
polynomial $\Delta J^0$ to $G_\psi$ we get the results for $G$ defined
with the covariant current (e.g.
\cite{tsu},\cite{hoso},\cite{banerjee}). 
                                 
 It is also possible to relate the Schwinger terms of
$G_{A/\psi}$ obtained in the present paper to
those in \cite{schwiebert},\cite{shizu}. Although the
consistent current was used  in \cite{schwiebert},\cite{shizu} and the
algebra of $G$ coincides with
\eq{result}, the Schwinger terms in the algebra of $G_{A/\psi}$ are
different. This can be explained by studying the time evolution
operator $U$. If we assume that  
\bm{rcl}
U(x_0,x_0) \& =\& \id \label{free2}\eg 
in \eq{free} we would arrive at the results of \cite{schwiebert}. In
\cite{schwiebert} \eq{free2} was used implicitely, since the commutators of
$G_{A/\psi}$ were calculated as defined in \eq{free}. 
Condition \eq{free2} is equivalent to gauge invariance of 
the underlying theory. In \cite{shizu} such a theory was constructed 
by adding a Wess-Zumino field. In this theory $G$ is not the full Gauss law
operator, since $G$ does not generate gauge transformations on the
Wess-Zumino field. The results in \cite{schwiebert} coincide
with those of \cite{shizu}, which confirms the interpretation.   
                 
The double commutators have been derived by using the
integrability of $W[A]$. Double commutators, which 
contain at least one $G$ agree with the iterative results. This 
indicates, that $G$ is a well defined operator on quantum level. The
Jacobi identity is satisfied by our results (eqs.
(\ref{nichtit}),(\ref{ueber3}),(\ref{ueber4}) and (\ref{currents}).
This is an outcome of the integrability (consistency) of $W[A]$, since the
double commutators are related to derivatives of $W[A]$ with respect to
the gauge field. In an iterative calculation one drops terms ensuring
the integrability (see \eq{double1} and the following discussion),
which leads to the violation of the Jacobi identity. 
Since double commutators containing at least one $G$ agree
with the iterative result, there is only one non-iterative term (up to
a minus sign)
\bg
\& \frac{1}{24\pi^2}\epsilon^{ijk}D^{ad}_i(x)D^{be}_j(y)
D^{cf}_k(z)\Tr\left[t^d\{t^e,t^f\}\right]\delta(\xv-\yv)\delta(\yv-\zv)\& \label{ddd}\eg
This can be traced back to the fact, that the gauge group
representation is projective and is related directly 
to the Schwinger term in the commutator $\left[G_A^a,G_A^b\right]$.

The non-iterative terms of double commutators, which are related via
cyclic permutation of the operators, are identical (no relative minus sign) (see
\eq{ueber3}-\eq{currents}). The BJL-type prescription 
of \cite{rothe} only works in this case. Therefore the results here
confirm the validity of the prescription in the case of $G_{A/\psi}$. 
Moreover, the connection of the non-iterative term \eq{ddd} 
to the integrability of the generating functional $W[A]$
indicates that this should hold for arbitrary double commutators.  

\subsection*{Acknowledgments}
I would like to thank H.~J. Rothe for helpful discussions.

\end{document}